\documentclass[conference]{IEEEtran}
\IEEEoverridecommandlockouts
\usepackage{cite}
\usepackage{amsmath,amssymb,amsfonts}
\usepackage{algorithmic}
\usepackage{graphicx}
\usepackage{textcomp}
\usepackage{xcolor}
\usepackage{nameref}
\usepackage{listings}
\def\BibTeX{{\rm B\kern-.05em{\sc i\kern-.025em b}\kern-.08em
    T\kern-.1667em\lower.7ex\hbox{E}\kern-.125emX}}
    
\usepackage{acronym}
\newacro{ips}[IPS]{Intrusion Prevention System}
\newacro{ids}[IDS]{Intrusion Detection System}
\newacro{yaas}[YAAS]{Yet Another Attribution Scheme}
\newacro{gamfis}[GAMfIS]{}
\newacro{ioc}[IoC]{Indicator of Compromise}
\acrodefplural{ioc}[IoC]{Indicators of Compromise}

\begin{document}

\title{Secure (S)Hell: Introducing an SSH Deception Proxy Framework}

\author{\IEEEauthorblockN{1\textsuperscript{st} Daniel Reti}
\IEEEauthorblockA{Intelligent Networks Research Group\\
German Research Center for Artificial Intelligence\\
67655 Kaiserslautern, Germany\\
Email: }
\and\IEEEauthorblockN{2\textsuperscript{nd} David Klaaßen}
\IEEEauthorblockA{Intelligent Networks Research Group\\
German Research Center for Artificial Intelligence\\
67655 Kaiserslautern, Germany\\
Email: david.klaassen@dfki.de}
\and
\IEEEauthorblockN{3\textsuperscript{rd} Hans Dieter Schotten}
\IEEEauthorblockA{Intelligent Networks Research Group\\
German Research Center for Artificial Intelligence\\
67655 Kaiserslautern, Germany\\
Email: schotten@dfki.de}
}

\author{\IEEEauthorblockN{Daniel Reti\IEEEauthorrefmark{1},
David Klaaßen\IEEEauthorrefmark{1}, Simon Duque Anton\IEEEauthorrefmark{1}\IEEEauthorrefmark{2}, Hans Dieter Schotten\IEEEauthorrefmark{1}\IEEEauthorrefmark{2}}
\IEEEauthorblockA{\IEEEauthorrefmark{1}Intelligent Networks Research Group, 
German Research Center for Artificial Intelligence, Kaiserslautern\\}
\IEEEauthorblockA{\IEEEauthorrefmark{2}Chair for Wireless Communication and Navigation, 
Technische Universität Kaiserslautern\\
Email: {firstname}.{lastname}@dfki.de}}

\maketitle

\begin{abstract}
Deceiving an attacker in the network security domain is a well established approach, mainly achieved through deployment of honeypots consisting of open network ports with the sole purpose of raising an alert on a connection. With attackers becoming more careful to avoid honeypots, other decoy elements on real host systems continue to create uncertainty for attackers.
This uncertainty makes an attack more difficult,
as an attacker cannot be sure whether the system does contain deceptive elements or not.
Consequently,
each action of an attacker could lead to the discovery.
In this paper a framework is proposed for placing decoy elements through an SSH proxy, allowing to deploy decoy elements on-the-fly without the need for a modification of the protected host system.
\end{abstract}


\section{Introduction}
Security of productive IT-systems is never complete.
Despite employing best practices and up-to-date \ac{ids} and \ac{ips} tools,
attacks will continue to be executed successfully.
Best practices such as segmentation,
strict firewall and traffic rules make it harder for an attacker and are capable of eliminating the majority of automated attempts to attack a system.
However,
a focused attack is harder to detect and mitigate.
A first step for an organisation to take countermeasures against an attack is obtaining knowledge that there is an ongoing attack.
The average time of detection of a data breach is 280 days,
according to a study performed by IBM in 2020~\cite{IBMSec.2020}.
Thus,
detection is an important,
yet difficult first step in mitigation of an attack.
Deception technologies can be employed as sensors with relatively low false positives and exceptional detection capabilities.
The most common deception technique are honeypots which simulate a host on the network with the purpose to raise an alert on interaction while wasting the attacker's resources. 
Ideally,
an attacker cannot distinguish between a productive and a deceptive resource.
If any interaction with the deceptive resource occurs,
the attacker's presence is made aware to the operators,
while at the same time collecting information about the attacker's goals,
methods and tools.
Such resources are a valuable tool in IT-security.
In this work,
the introduction of deceptive techniques into SSH-communication is presented.
A proxy server is used to inject deceptive elements into the SSH communication.
While a valid user solely interacts with productive elements,
an attacker is likely to be lured to interact with interesting-sounding elements,
which in turn alerts operators of a malicious activity.
The contributions of this paper are the following:
\begin{itemize}
  \item Give a literature review on the usage of SSH Proxies and Deception Proxies
  \item Introduce an SSH deception proxy to inject decoy elements into SSH traffic 
  \item Give insights from the writing a PoC SSH Deception Proxy based on Python Paramiko 
  \item Discuss the implementation on a detailed level
\end{itemize}
% Simon Says: Might be just me, but I'd not make the section names bold, but use references instead
The remainder of this work in structured as follows: In Section \ref{background} decoy elements and deception are introduced. In Section \ref{architecture} an attacker model is given and used to explain the benefits of the proposed concept. In Section \ref{implementation} is explained how the proposed concept was implemented. In Section \ref{discussion} the benefits and difficulties are discussed and in the conclusion as short summery of the work and the results is given.

\section{Background}\label{background}

\subsection{Deception Technology}
Deception Technology, which is also referred to by the term Cyber Deception, describes the application of fake resources and obfuscation in IT environments \cite{FraunholzDemystify}. A popular example is a honeypot which typically simulates open TCP ports and raises an alert upon connection.
The typical goals of deception are to learn about attack techniques, waste attackers resources and time, create uncertainty and detect malicious activity while distracting from real targets. The most common taxonomy by Whaley et al. distinguishes the deception categories simulation and dissimulation,
where dissimulation can be further described by the three categories masking, repacking and dazzling, while simulation is categorised into mimicking, inventing and decoying \cite{Whaley.1982}. 
Traditionally, deception is commonly applied by honeypots or as so-called honeytokens, which are lures and fake resources on a system. In this work, an SSH proxy is introduced that injects these kinds of lures and fake information into the shell responses of a connected host system. This allows to convert every Linux based system with SSH connectivity to be turned into a high-interaction honeypot and to plant honeytokens without the need to modify the system with a minimal configuration overhead.

\subsection{Secure Shells in Deception}
SSH is short for Secure SHell and describes a network protocol which is typically used for remote, network based command line interfacing with a machine. In contrast to its predecessor Telnet, SSH supports several encryption protocols, protecting it from man-in-the-middle attacks. By the time of writing, the search engine Shodan lists over 18 million devices
connected to the Internet with the SSH default TCP port 22 open and over 16 million devices stating using the OpenSSH implementation in the SSH welcome banner. As this service allows to remotely control a machine, connecting over SSH is a common target for attackers. Therefore, a host with an open, internet-facing SSH port is constantly visited by bots trying to connect with default credentials in order to acquire further systems to add to botnets. For this reason, popular honeypots such as cowrie simulate a host with an open and often weakly secured SSH
port. In order to manage the connection between clients and honeypots, SSH proxies are already available, such as HonSSH \cite{ThomasNicholson.2016} or the Cowrie Proxy Mode \cite{cowrie.}. This work proposes an SSH proxy as shown in Figure \ref{fig.DeceptionProxy}, which injects false information into and obfuscates real information in shell responses while maintaining
legitimate interaction with the host system for legitimate users. Administrators seeking information concealed by the proxy should know about the deception and thus not trust it.

\subsection{Attacker Model}
Attacker models are used to describe and classify an attacker in terms of knowledge,
resources,
and objectives.
Fraunholz et al. present a methodology %called \ac{yaas} 
consisting of a capability and threat rating,
which in turn are based upon skill and resources,
as well as motivation and intention respectively\cite{Fraunholz.2017f}.
However,
they discuss that especially intention and motivation are difficult to infer from \acp{ioc}.
Deception technologies,
especially honeypots,
are capable of monitoring activities and allow for the inference of an objective of the attacker.
Still,
there is guesswork involved.
Apart from attacker models that focus on the attacker,
attack models that are used to describe the attack,
rather than the attacker, 
are employed.
Famous attack models are and the Lockheed Martin Cyber Kill Chain~\cite{cyberkillchain} and the MITRE ATT\&CK matrix~\cite{mitre} which is based on the Cyber Kill Chain.
Both models structure the behaviour of an attacker,
regardless of objective,
resources or skill level.
Generally,
an attacker first has to breach the perimeter,
usually after reconnaissance and development of a suitable attack,
and then move laterally towards the target,
while establishing persistence,
escalating privileges and executing actions on the infected machines.
An attacker that is detected by an SSH-proxy will assume to have obtained initial access to a machine and will consequently try to establish persistence,
execute actions,
escalate privileges,
or, 
if the attacker's motive is curiosity,
exfiltrate data that is available.
All of these options can be mimicked by the proxy,
so the objective of an attacker can be derived from the actions performed.
Data bases,
personal files as well as opportunities for exfiltration can be provided given an attacker tries to steal data.
Kernel information, 
program versions and availability of credentials,
configuration files and programs can deceive an attacker to perform privilege escalation attacks given a persistent or higher privileged access is the objective.

%%Einfallstor SSH port + Gestohlener schlüssel oder Passwort Bruteforce.
%%Danach (in Bezug Setzen auf MITRE + Cyber Kill Chain): 
%Ziele:
%- Exfiltration
%    - Datenbanken, Persönliche Dokumente, Exfiltreation channels
%- Priv Esc / Persistance
%    - Enumeration (Welcher Kernel, Welche Programme (versionen), find credentials, find configuration files)
    
%Goal: Vehaltensmumster ableiten und deception Regeln schreiben
%In diesem Paper: erstmal high level: cat, ls überschreiben...

\subsection{Proxies for Deception}
The first publications to describe reverse proxies for deception were
published in 2017 independently by Han et al. \cite{Han.10302017} and by Ishikawa et al. \cite{Ishikawa.01052017}. Theses proxies inject additional cookies, forms, and HTTP parameters into HTTP traffic. A similar concept was described in 2018 independently by Fraunholz et al. and Papalitsas et al. \cite{Fraunholz.10152018} \cite{Papalitsas.2018}. Both works describe a deception reverse proxy altering HTTP and other application protocols by hooking packets. Han et al.'s proxy is based on hoxy, the proxy presented by Fraunholz et al. is based on mitmproxy, both being HTTP sniffing and manipulation libraries for debugging purposes. The proxy developed by Papalitsas et al. is dissecting HTTP packets with a self-written plugin and is limited to structured data such as JSON and LDAP. To the authors' best knowledge no publication describes the manipulation of SSH traffic trough a proxy.
Other uses of proxy servers in the context of deception exist, for example honeypots that are based on an open proxy server to attract spammers \cite{Mushtakov.29.01.201801.02.2018} or SSH proxies that are used to forward traffic to a honeypot like the cowrie proxy mode.

\section{Architecture}\label{architecture}
\begin{figure}
\centering
\includegraphics[scale=0.9]{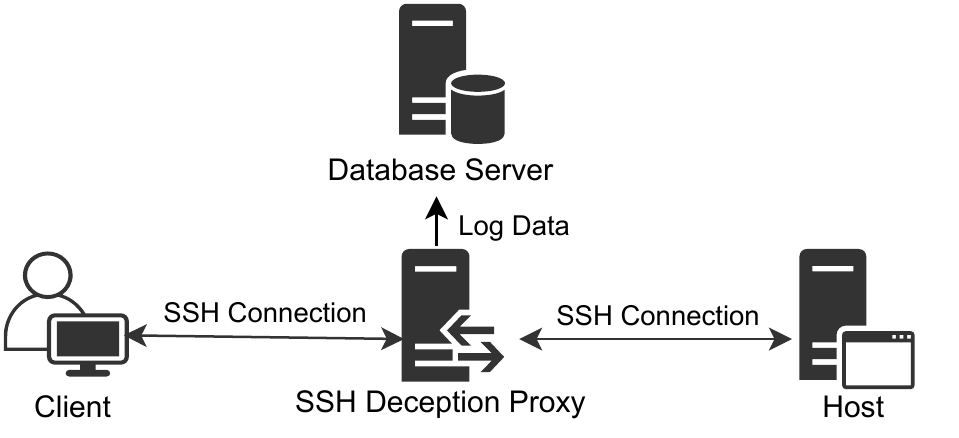} \\
\caption{Traffic flow of the proposed SSH deception Proxy}
\label{fig.DeceptionProxy}
\end{figure}

In this chapter the overall architecture of the proposed SSH deception proxy is described.
Each required architectural characteristic is presented in the following subsections.
As shown in Figure~\ref{fig.DeceptionProxy}, the proxy is placed as a reverse proxy, i.e. in front of the host with every client connection having to pass the proxy.
% passing traffic?
The traffic passed through the proxy is logged and modified by adding honey tokens into the shell responses. A detailed depiction on the design decisions
is given in the remainder of this chapter.
\iffalse
- Requirements: Host muss nicht modifiziert werden, Ease of Deployment (z.B. möglichst wenig konfigurationsaufwand und eingriff auf den Host), Modularer aufbau für neue Regeln, Möglichst wenig interferenz mit legitimer Nutzung des Hosts, Angreifer soll nicht merken dass er mit einem Deception Proxy interagiert -> standard operationen müssen für fake files unterstützt werden
- A proxy is used for ease of deployment and no modification of host
- How does a transparent reverse proxy operate? Packet flow Client -> Proxy -> Host -> Proxy -> Client
- Deception Tokes: add fake files and manipulate file contents existing files, since in linux everything is a file, all system configuration, non-encrypted information and credentials can be faked. Eg. fake a kernel version with cat /proc/version, add fake credentials in /etc/passwd, 
- For this the output of specific commands has to be altered, e.g. cat, ls, piping operators, but can also be used other binaries like uname
- To avoid easy detection through banner grabbing the SSH welcome banner of the original host is copied to the proxy.
- SSH proxy with preshared keys and configurable deception rules
- Suspicious commands can be logged and raise alerts
\fi

\subsection{Design Guidelines}
% Requirements: Host muss nicht modifiziert werden, Ease of Deployment (z.B. möglichst wenig konfigurationsaufwand und eingriff auf den Host), Modularer aufbau für neue Regeln, Möglichst wenig interferenz mit legitimer Nutzung des Hosts, Angreifer soll nicht merken dass er mit einem Deception Proxy interagiert -> standard operationen müssen für fake files unterstützt werden
These guidelines describe a desired solution. While most of these guidelines will be followed,
a trade-off between implementation effort and fulfillment of requirements is made for the actual implementation. This in turn affects the architecture and the reasons for these decision will be explained.
\subsubsection{Modification of Host}
It is desireable for the host to be unaffected by the proxy. Some changes are unavoidable, such as log files on the host no longer representing the actual source of connections made to it as all connections to the host are made through the proxy.
\subsubsection{Ease of Deployment}
The proxy should require as little configuration as possible. However, since SSH represents an encrypted stream, knowledge about the host is required in order to correctly interpret the stream in real time. Common problems are: 
\begin{itemize}
\item	recognizing the beginning of a command
\item	recognizing the end of command output
\item 	providing file interaction for decoy files
\item 	dealing with scripts
\end{itemize}
Solutions for these problems will be discussed in detail in Section~\ref{sec:implementation}. %\autoref{Implementation} does this work? apparently not
\subsubsection{Modularity for new rules}
Rules concerning the manipulation of commands should be easy to implement and follow a common scheme.
\subsubsection{Interference with Legitimate Use} % maybe combine this and the following guideline and the subsection "Transparent Reverse Proxy" into one guideline "Transparency"
Legitimate users should not be impacted by the proxy. As such, the proxy may only provide additional functionality or - where it is explicitly defined by the configuration set up by the administrator - block certain actions which are not required by legitimate users. \\
% Versteh' ick nüsch.
%This feature mainly shows how easily reverse proxies can be extended for additional functionality. % ?
\subsubsection{Transparency for Attackers}
Attackers should not notice any interference by the reverse proxy. Any additional functionality implemented by the proxy, such as adding files to the output of \texttt{ls}, must be accompanied by features which complete the illusion, e.g.
simulate the presence of these files on the host file system. These decoy files are saved on the proxy and whenever a command sent to the host by the client requires any of these files, features must be provided to implement this functionality. It is assumed that any interaction with decoy files is malevolent.

\subsection{Transparent Reverse Proxy}
% How does a transparent reverse proxy operate? Packet flow Client -> Proxy -> Host -> Proxy -> Client
A normal proxy  is invisible to the host. A client connects with the proxy and it forwards the packets in both directions. The host is not aware that it communicates with a proxy.\\
Reverse proxies are invisible to the client. The client connects with the reverse proxy, mistaking it to be the host. Since it is a transparent proxy, the client does not notice any difference to communicating with the normal host.\\
SSH adds several complications.

% A proxy is used for ease of deployment and no modification of host

\subsection{Command and Output Manipulation}
% Hier fehlt mir irgendwie ein Statement dazu, wie mit der Verschlüsselung umgegangen wird, was ja auch im Vorhinein als Problem identifiziert wurde
% For this the output of specific commands has to be altered, e.g. cat, ls, piping operators, but can also be used other binaries like uname
The manipulation of the SSH stream is the main feature of the proxy. 
%However, since this is the only way the client interacts with the host (excluding side-channel attacks) everything can be manipulated. 
%This feature is restricted by the transparency guideline and since most functionality should not be restricted, most vulnerabilities will continue to exist. \\
What this feature can do is recognize specific commands such as \texttt{uname -s} and instead of returning the true kernel name, as would be expected, return a fake kernel name which invited the attacker to try exploits the actual kernel is not vulnerable to. //
While manipulation of the stream is powerful enough to manipulate the attacker's perception of the host system completely, legitimate use would be impeded. E.g. for some hosts their legitimate users may require information about the kernel's version number, in order to accomplish their tasks. A trade-off is made.\\
% Suspicious commands can be logged and raise alerts
It is also possible to catch commands that are deemed suspicious and, as described earlier, ban or deceive the responsible IP address.

\subsection{Deception Tokens}
% Deception Tokes: add fake files and manipulate file contents existing files, since in linux everything is a file, all system configuration, non-encrypted information and credentials can be faked. E.g. fake a kernel version with cat /proc/version, add fake credentials in /etc/passwd, 
Fake files should seem to be part of the host's file system. This lends itself readily?
for several features:
\paragraph{Honey Files} Honey files can be added to command output. For example fake credentials in \texttt{~/passwords.txt} can be added. When using these credentials to connect to the host, the proxy catches their usage and can
stop the IP from reconnecting, redirect the connection into a sandbox, or implement other appropriate countermeasures.

\paragraph{Overwriting critical information contained in files} Some files, such as \texttt{/proc/version}, contain critical information which can reveal weaknesses about the host. These files can be either overwritten with wrong information, making the attacker waste time on exploiting vulnerabilities fixed in previous versions or their content can be hidden completely. Since "everything is a file" in Linux, most configuration information can be faked.

This feature implies the implementation of other features to avoid easy detection. Pipes and programs on the host may be ordered to use fake files. While pipes can be implemented easily enough, programs on the host require either the fake file to be copied from the proxy to the host, or detection is unavoidable. Even in these cases a generic error message could replace the program's error message in order to deceive the attacker.

\subsection{Bannergrabbing on the SSH proxy}
% Finde die Argumentation schwierig. Banner werden modifiziert -> das kann man als Angreifer erkennen -> das kann man als Verteidiger verhindern, indem man Banner nicht modifziert oder sie modifiziert. 
% To avoid easy detection through banner grabbing the SSH welcome banner of the original host is copied to the proxy.
Bannergrabbing from the host is caught by the proxy. This may lead to easy detection if the returned banner does not fit the features provided. There are two ways to deal with this. The first is to simply return the banner of the host, the second is return a modified banner. Returning the true banner will avoid detection, but e.g. lowering the minor version number may lead to wasting time on already patched vulnerabilities, though that is unlikely with SSH.

% SSH proxy with preshared keys and configurable deception rules 
%
% I don't understand.
%

\section{Implementation}\label{implementation}
\label{sec:implementation}
\iffalse
- Softwarestack beschreiben: Python Paramiko\cite{Paramiko}, Linux host, 
- Paramiko is used as Server and as a client simultaniously
- On incoming connections, the Server starts a handler thread with a client to connect to the host
- Handler funktionsablauf
- What was modified in Paramiko (Password checking, banner)
- Difficulities: catch commands in interactive shells like VIM, check for promt
    know and interpret when a command is finished executing (also check for promt)
- How are the honeyfiles handled.
    Local files on proxy overwrite remote files
- Regex is used to catch commands
- Man console codes \cite{LinuxKernel.}
\fi

\subsection{Basic Structure}
% Softwarestack beschreiben: Python Paramiko, Linux host, 
% Paramiko is used as Server and as a client simultaniously
% On incoming connections, the Server starts a handler thread with a client to connect to the host
This implementation runs on Linux and is implemented in Python3. The python package Paramiko \cite{Paramiko} implements SSHv2 and is used to create both an SSH server for the client to connect to and an SSH client in order to connect to the host. \\
When a client connects to the host, a \texttt{handler} function is called in a new thread and passed the raw socket as a parameter. The \texttt{handler} then creates the adapted Paramiko server object and checks whether the credentials and the IP are eligible for a connection. \\
When the handler receives a command, the appropriate \texttt{command handler} is chosen and, if necessary, 
% würde im passiv bleiben: the output is modified accordingly
the output modified accordingly.
The configuration is done via a \texttt{Config.py} file which contains a python dictionary. 
In the following, \texttt{config} refers to this dictionary, \texttt{handler} refers to the handler function invoked in the main thread and \texttt{command handler} refers to functions modifying the output according to the command they were created for. \\
Output is the stream from the host to client forwarded through the proxy. It refers to the output that specific commands trigger. \\

\subsection{Paramiko Server}
% What was modified in Paramiko (Password checking, banner)
This Paramiko server had to be customized to provide banner grabbing and credential checking.
\subsubsection{Bannergrabbing}
Since this is the server the client connects to, its banner is the one the client receives when grabbing for banners. The \texttt{config} currently contains the banner for OpenSSH 7.9p1 running on Debian and this banner is sent.
\subsubsection{Credentials}
The credentials are checked by trying to create a connection to the host with the given credentials. Should this not be possible, the credentials are wrong and this is returned. In the case that the credentials are honey tokens in form of fake credentials, as described above, the login attempt can be assumed to be adversarial.
%zu spezifisch:
%the connection is terminated and the IP address is added to the list of blocked IP addresses as it is assumed to be an attacker. The IP address is considered to be no longer eligible for a connection. Checking whether the IP address is eligible for a connection is not done here, the handler checks this before creating the server object. For ease of implementation, only password authentication is currently enabled.

\subsection{\texttt{handler} function}
% Handler funktionsablauf
\begin{figure}[]

\centering
\includegraphics[scale=0.8]{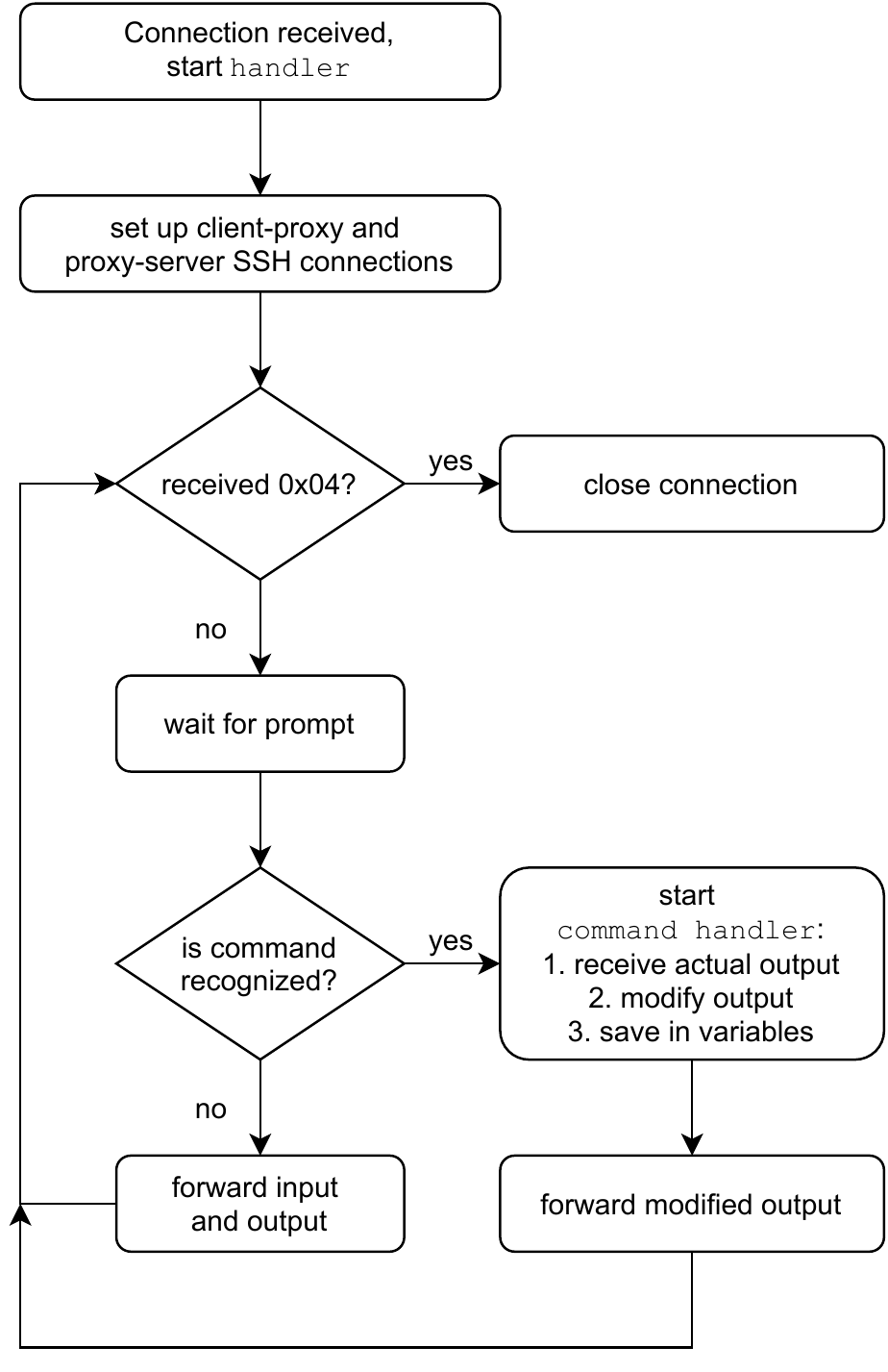} \\
\caption{Overview of the Control and Data Flow in the \texttt{handler}}
\label{fig:handler}
\end{figure}
The \texttt{handler} function, as depicted in Figure \ref{fig:handler}, is responsible for dealing with an individual connection. It is called with a raw socket and checks whether the IP address is eligible for connection. If yes, a Paramiko server object is created in order to deal with the SSH connection. The credentials the client entered are used to authenticate to the host. \\
Usually, SSH servers will now send a MOTD ("message of the day") which may contain sensitive information such as the last time a user logged in. This is forwarded in order to avoid detection. \\

The handler-function is responsible for the following three tasks in the communication.

\subsubsection{Recognizing the end of command output}
One of the biggest challenges in dealing with a bidirectional stream is that there is no way of knowing for sure that command has finished executing. While no good solution was found, the best working solution is to wait for a prompt. \\
Waiting for the command output is necessariy in the case that a \texttt{command handler} recognizes the command because the manipulation of many commands requires the actual output. E.g. the output of \texttt{ls} is sorted and thus the output has to be known in order to not be immediately obvious to any attacker by appending the fake file names to the output. 

\subsubsection{Recognizing a prompt}
% Regex is used to catch commands
The typical prompt has the form: \\
\texttt{username@host:path\$ } \\
A regular expression is used to check whether the current buffer of received symbols ends with a prompt. Since this prompt can have several forms, knowledge about the host system is necessary. E.g. the prompt could end with a \texttt{\#}, or take an altogether different from. \\
This form of output recognition has the weakness that whenever a prompt is displayed, e.g. due to being entered into a text editor, the current buffer ends with a prompt. Thus, a new command is expected, causing difficulties and commands may evade the \texttt{command handler} functions as they are treated as output rather than commands. \\

\subsubsection{Unrecognized commands}
% Difficulities: catch commands in interactive shells like VIM, check for promt
%    know and interpret when a command is finished executing (also check for promt)
Recognized commands will be sent to the appropriate \texttt{command handler} which is described below.  Unrecognized commands cannot be dealt with by waiting for the correct output and then sending it because it is not known whether the command opens an interactive program such as VIM or not. Interactive programs require the bidirectional stream to be forwarded into both directions. This is done until the forwarded characters end with a prompt, as it announces that the interactive program has finished executing and the terminal is waiting for a new command to be sent. \\
Difficulties arise if unrecognized commands are called with decoy files as arguments. Two solutions present themselves and were discussed earlier.
% Firstly, it is possible to send a command to the server, that creates the files, then wait for the command to finish and send a command to delete the files. This however interferes with the server and discovery is risked due to decoy

\subsection{\texttt{command handler} Functions}
The correct \texttt{command handler} is chosen according to the command it is supposed to deal with. In the current implementation this is done with \texttt{if, elif, else}, where \texttt{if} and \texttt{elif} recognize commands and \texttt{else} deals with unrecognized commands as described above. \\

\subsubsection{\texttt{command handler} structure}
Modularity is desired and as such there is a common scheme for all \texttt{command handler} functions:
\begin{enumerate}
\item 	receive actual command output
\item	modify command output according to rules
\item 	set \texttt{send\_before\_modified\_response, modified\_response, send\_after\_modified\_response}. This allows for chaining command handler should it be desirable. E.g. the command \texttt{head} can be implemented by calling the handler for \texttt{cat} and then removing every line in \texttt{modified\_response} after the first ten lines.
\end{enumerate}

\subsubsection{Honey File Injection}
% How are the honeyfiles handled.
%   Local files on proxy overwrite remote files
Currently, only \texttt{command handler}s for dealing with \texttt{ls, cat} and pipes are implemented. As such, basic functionality for decoy files is implemented as a proof of concept. \\
The honey files on the proxy are organized by providing the \texttt{config} with a path to a folder that serves as the root for a decoy file system. The files contained therein are overlaid onto the real files on the host. Decoy files take priority over real files in order to be able to overwrite them. \\
Another solution would be to have two file system roots in the \texttt{config}: one for hiding files, e.g. by deleting entries from \texttt{ls}, and another root for adding decoy files which do not overwrite files. This would be a better solution as overwriting files can interfere with legitimate use. Currently, file names have to be chosen with care. \\
While it is possible to create entire honey directories, the current implementation does not support that feature. \\
Most shells feature \texttt{~} as a character that is short for the current user's home directory. Since this home directory is not immediately obvious to the stream, \texttt{pwd} is called in the beginning to be able to translate absolute file paths containing \texttt{~} into the proxy's file system path. Another solution would be choosing \texttt{/home/(username)/} as a heuristic.

\subsection{Linux Terminal special commands}
% Man console codes \cite{LinuxKernel.}
% dimensions of the display
The stream of characters encoded by SSH is parsed by (in this case) Linux terminals. Linux terminals have several special characters \cite{LinuxKernel.} which require work to be parsed into a command correctly. E.g. backspace (\texttt{0x07}) deletes a character and moves the cursor to the left. Without parsing this, any command with a backspace would not be recognized correctly.\\
\texttt{passs0x07wd} $\neq$ \texttt{passwd}
\subsubsection{Current Implementation}
If dealing with single line inputs, it is sufficient to project the 2D Linux terminal into a single line. The current implementation uses two arrays of characters, one to the left and one to the right of the cursor position and each special character modifies these arrays accordingly. This deals with special characters related to cursor position as well as normal characters by appending them to the left array.
\subsubsection{Other Special Characters}
Some special characters, e.g. color commands, have no effect on the cursor position or the buffers. However, when sorting \texttt{ls} output according to files names, both removing color from the output as well as coloring added honey files/executables and restoring the former color to all entries is necessary.

\subsection{Command History and TAB Autocompletion}
Shells have several features which are commonly used and as such need to be implemented in order to not be immediately obvious to the user.

\subsubsection{Command History}
When pressing the UP key, previously entered commands are shown. If executing own commands, they are usually entered into the command history. If the implementation is keeping track of the command history, it is possible to "skip" own commands by sending an additional UP or DOWN key to the host. \\
In the current implementation a SPACE is prepended to the command and on the current host system this prevents commands from being entered into the command history. This may not be possible on other devices if no additional configuration effort is undertaken.

\subsubsection{TAB Autocompletion}
Autocompletion should work on both honey and real files. As autocompletion results are usually alphabetically sorted it is possible to insert honey files at the right location and avoid detection.

\subsection{PostgreSQL logging}
%psycopg2
The current implementation supports logging via Psycopg2's\cite{Psycopg2} logging module offers the possibility to extend \texttt{logging.Handler} into subclasses. The implementation makes use of this.
Database address, name, username and password are saved in the \texttt{config}. \\
The implementation tracks suspicious commands, such as usage of honey files, and sends the suspected attacker's IP, username and the transgression to the database. Other events could also be tracked, e.g. suspiciously high amounts of logins from one IP address, blocking of IPs, etc.\\

\section{Discussion}\label{discussion}
%- Setting an alias for commands allows to bypass the proxy
%- Using commands without specified rules allows to bypass the proxy

The SSH deception proxy is a promising approach and has been shown to be feasible in a proof of concept implementation. Not needing to modify the host and the simple deployment are big advantages as well as the possibility to log commands. Still, there are drawbacks regarding the detectability by adversaries and possibilities to bypass the proxy. The problem is that only a limited set of commands is being caught by the proxy, therefore using commands that are not in set set allows to bypass the proxy and eventually notice the proxy. For example listing all files in a directory with \texttt{ls} will have all honey files injected by the proxy, while \texttt{echo *} will also list all files while bypassing the proxy. Another possible way to bypass the proxy could be the use of aliases or creating another remote connection from the host that bypasses the proxy. Additional filters and restrictions would have to be put in place to restrict these bypasses. 

Nonetheless the drawbacks considered bearable, as in the worst case the security of the system only falls back to the state of having no deception proxy is in place, therefore the proxy could still be considered a security improvement when only a small number of attackers are successfully deceived. Therefore it might be possible for careful attackers to recognize and bypass the deception in place, but it still has value for detecting all other adversaries.
\section{Conclusion}
In this paper an SSH based deception proxy was proposed and a proof of concept implementation was described. As deception is a promising technique for attacker detection and distraction, the SSH deception proxy allows to combine these features of deception with the advantages of a proxy, which are ease of deployed and replacing the need for modifying the host. Although a bypass of the deception proxy is theoretically possible, it would only reduce the security of the protected host to the state it was before the deception proxy was in place. Future work includes identifying and evaluating the best deception techniques that could be used using the SSH deception proxy.

\section{Acknowledgement}
This research was supported by the German Federal Ministry of Education and Research (BMBF) within the IUNO Insec project under grant number 16KIS0932, and within the SCRATCh project under grant number 01IS18062E. The SCRATCh project is part of the ITEA 3 cluster of the European research program EUREKA. The responsibility for this publication lies with the authors.
\bibliographystyle{IEEEtran}
\bibliography{literature}
\vspace{12pt}

\end{document}